\documentstyle[aps]{revtex}
\def\A{\mbox{\bf A}}
\def\unitvec{\mbox{\boldmath $e$}}
\def\e{{\rm e}}
\def\E{\mbox{\bf E}}
\def\B{\mbox{\bf B}}
\def\v{\mbox{\bf v}}
\def\r{\mbox{\bf r}}
\def\f{\mbox{\bf f}}
\def\be{\begin{equation}}
\def\ee{\end{equation}}
\begin{document}
\input epsf
\draft
\twocolumn[\hsize\textwidth\columnwidth\hsize\csname
@twocolumnfalse\endcsname
\date{September 1999; revised November 1999}
\title{Thermal fluctuations in macroscopic quantum memory}
\author{S. Khlebnikov}
\address{
Department of Physics, Purdue University, West Lafayette, IN 47907, USA
}
\maketitle
\begin{abstract}
We describe macroscopic quantum memory devices based on type-II toroidal superconductors
and estimate in one case and compute in another the rates at which quantum
information stored in these devices ``degrades'' because of thermal fluctuations.
In the case when the entire solid torus is superconducting, the Boltzmann factor in
the rate corresponds to a well-defined critical fluctuation, and the rate
is suppressed exponentially with the linear size of the system. In the case when
superconductivity is confined to the surface of the torus, the rate is determined 
by diffusive motion  of vortices around the torus and does not depend exponentially
on the linear size; we find, however, that when the two dimensions of 
the torus are comparable the rate does not contain the usual volume enhancement 
factor, i.e. it does not grow with the total surface area of the sample. 
We describe a possible way to write to and read from this quantum memory.
\end{abstract}
\pacs{PACS: 85.25.Hv, 03.67.Lx \hfill PURD-TH-99-07 \hfill quant-ph/9909024}
\vskip2pc]

\section{Introduction}
Quantum memory is a device capable of reliably storing linear 
superpositions of quantum states.
It will be a part of quantum computer when (if) that latter is 
finally built and may be useful for other applications as well.
(For a recent review of quantum computing with an emphasis on fault
tolerance see ref. \cite{Preskill}.)

To work as quantum memory, a physical system must satisfy a number of 
requirements. First, it must have at least two fairly stable quantum states. 
These states form a basis for linear combinations that can be stored 
in the device. For example, the basis may be formed by perturbative quantum
states built near local energy minima, and stability of the basis states may 
be ensured by a large potential barrier separating them. 
In such cases we will
loosely refer to the basis states as the ground states, or vacua,
even though these ground states may not be degenerate in energy
and in some cases may contain localized excitations.
We note, though, that for some purposes it may be desirable to have ground
states that actually are (nearly) degenerate in energy. 
If two basis states $|\psi_1\rangle$ and $|\psi_2\rangle$ forming a linear 
combination
\begin{equation}
|\psi\rangle = c_1 |\psi_1\rangle + c_2 |\psi_2\rangle \; 
\label{gs}
\end{equation}
are degenerate, the ratio $c_2/c_1$ will be preserved by the evolution.
When the basis states are not degenerate, the relative magnitude of $c_1$ 
and $c_2$ will be preserved, but not the relative phase. 
The relative magnitude, however, can be arbitrary. In comparison,
a classical two-state system will only store two values, referred to as 0 and 1.

For long-time quantum storage, one will probably need to build in some
redundancy, so that the basis states refer to
many  microscopic (local) degrees of freedom. However, redundancy
is helpful in protecting quantum information only when the local degrees
of freedom in the basis states are sufficiently entangled, 
i.e. the basis states cannot be identified by local measurements.
This condition rules out, in particular, any system in which 
a ground state degeneracy is due merely to spontaneous symmetry breaking 
by a local order parameter. (To see why, consider an easy-axis magnet, 
in which magnetization can be in one of two directions. The direction 
of magnetization can be found by measuring local magnetization in a relatively 
small region.) 
The reason why entanglement of local degrees of freedom is
necessary for long-time quantum storage is that local measurements will
in effect be performed by external noise,
and if they can indeed distinguish between the basis states they will 
destroy the stored quantum information (the Schr\"odinger-cat scenario).

In a real sample, tunneling transitions between the basis states will 
cause quantum memory to deteriorate. Nevertheless, if
the basis states are sufficiently entangled, tunneling between them 
will have to involve 
many local degrees of freedom, and the tunneling probability 
will be strongly suppressed. Known examples \cite{Wen&Niu,Kitaev} include 
fractional quantum Hall and similar types
of rigid ground states on tori. 
In these cases, a typical tunneling fluctuation consists of creating a vortex-antivortex 
pair, transporting the vortex and the antivortex around the torus, along topologically
distinct paths, and then annihilating the pair.
It has been argued that, at zero temperature, the tunneling probability
(and the associated energy splitting between the ground states) is generically of 
order $\exp(-L/l)$ where $L$ is the size of the system, and $l$ is some
correlation length. So, the zero-temperature 
tunneling should not be a problem in practice, as long as one can keep the size of 
the system sufficiently large.

Of more concern are thermal fluctuations. At finite temperature, there will be a sea
of vortex-antivortex pairs, with density proportional to $\exp(-F_0/T)$, where $F_0$ is the
free energy of a single vortex. (This assumes that the temperature is still low enough, so
that no phase transition occurs.)  One expects that motion of these ``preexisting'' vortices
can effect transitions between ensembles built near different ground states. 
The question is, then, what is the rate
of such transitions, at a given temperature $T$. A single transition is sufficient
to destroy the stored quantum information.
So, the rate of the transitions will also be the rate at which the quantum 
information ``degrades'',
and the corresponding time will be an estimate for the maximal duration of reliable storage.

In this paper, we will compute in one case and estimate in another
the rates of finite-temperature transitions between ensembles corresponding to
different ground states for some of the simplest systems exhibiting multiple ground 
states and macroscopic
entanglement. One system is a type-II superconducting film grown on the surface of a torus.
In Sect. 2 we review the origin of multiple ground states in a type-II superconductor on a torus.
The presence of multiple ground states in this case
can be seen either via manipulations with vortices and single electrons, which
produce a nontrivial phase when transported around each other, or via a semiclassical argument. 
Transitions between different classical vacua are topological transitions, which change
a winding number of the gauge and Higgs fields. In Sect. 3 we construct
a correlator that measures the rate of topological transitions at finite temperature.
This correlator is analogous to the one proposed in ref. \cite{KS} to measure
the rate of topological transitions in the electroweak theory. In Sect. 4 we compute the rate.
The main ingredient of the computation is that vortices are well 
separated, and their motion is diffusive, i.e. associated with a large viscosity.

Our main result, for a film of fixed thickness,
is that although the rate of topological transitions is indeed proportional
to the vortex density, and so is not suppressed exponentially with $L$ 
(the size of the system), there is a power-law suppression. This suppression can be
described by saying that there is no volume enhancement of the rate, i.e. as long as
the two dimensions of the torus stay comparable, the rate will not grow 
with the total volume (while the total number of vortices of course 
will). Equivalently, the rate per
unit volume will decrease with the total volume. This absence of macroscopic enhancement
is directly related to the diffusive nature of the vortex motion.

It is easy to redesign the device so that the suppression of the 
finite-temperature rate becomes exponential with $L$. 
Imagine making the  superconducting film thicker, so that vortices resolve into 
Abrikosov flux lines; the free energy of those grows linearly 
with their length. In the limiting case, which is the second system we consider,
the entire solid torus is superconducting, and
a topological transition is mediated by a well defined critical 
fluctuation---a critical flux line, whose energy is proportional to $L$.
The Boltzmann factor in the rate is 
$\exp(-E_0/T)$, where now $E_0 \propto L$, so the finite-temperature rate 
is suppressed exponentially with $L$.
The zero-temperature tunneling rate is suppressed even stronger, as an exponential
of $L^2$. So, a solid superconducting torus (or a wire, or a ring, or a hollow cylinder)
is a good candidate for stable quantum memory. In the concluding section we discuss 
a possible way of writing quantum information to and reading it from this device.

We nevertheless retain interest in the two-dimensional case (the film), because
a universal quantum computation 
is theoretically possible with non-Abelian anyons \cite{Kitaev}, and 
systems in which those have been argued to occur \cite{nonab} are 
two-dimensional. In the concluding section we also discuss whether our 
results teach us anything about these more complex cases.

\section{Ground states of a toroidal superconductor}
Existence of multiple ground states in
a type-II superconductor on a torus can be deduced
from the presence of two types of local excitations, vortices and 
single electrons, with their corresponding values of flux and charge.
It can also be obtained from an explicit semiclassical construction of the
ground states. In this section, we use the Ginzburg-Landau (GL) theory for 
description of the ground states. 
We interpret the GL expression for energy as an effective Hamiltonian for
slow degrees of freedom (rather than as a thermodynamic potential, 
like free energy). So, we treat the GL fields as quantum fields. 

The GL Hamiltonian of a superconductor is
\begin{equation}
H = \int d^{3} x \left( \zeta |(\nabla + ig\A)\psi|^{2}
- a |\psi|^{2} + b |\psi|^{4} \right) + H_{\rm EM} \; ,
\label{H}
\end{equation}
where $\zeta$, $a$, and $b$ are positive coefficients,
\begin{equation}
 g = 2e/c \; ,
\label{g=2e}
\end{equation}
$2e$ is minus the electric charge of a Cooper pair ($e>0$), 
and $c$ is the speed of light; $\hbar =1$ everywhere.
We concentrate on the extreme type-II case; the corresponding condition on 
the parameters is
\be
g^2 \zeta^2 \ll b  \; .
\label{typeII}
\ee
The Hamiltonian of electromagnetic field is taken, for simplicity,
in the relativistic form:
\begin{equation}
 H_{\rm EM} = {1\over 8\pi} \int d^{3} x \left( \E^{2} + \B^{2} \right) \; .
 \label{HEM}
 \end{equation}
In (\ref{H}), (\ref{HEM}) $\psi$ is the complex ``order parameter'' 
field (it is not really an order parameter because it is not gauge-invariant \cite{op}),
$\A$ is the electromagnetic
vector potential, $\E$ and $\B$ are the electric and magnetic fields. 

We first consider a superconducting film that extends from $z=0$ to $z=d$ in 
the $z$ direction and is periodic (toric) in the $x$ and $y$ directions.
These periodic boundary conditions define what may be called a ``mathematical''
torus, as distinct from the surface of a physical torus, a ``doughnut'', that one
may produce in a laboratory. Later, we will discuss the distinction in more detail
and will also consider the case when the entire solid torus is superconducting.

The vortex of the theory (\ref{H}) is a 
short (of length $d$)
Abrikosov flux line whose axis is parallel to the $z$ axis. A vortex
carries magnetic flux of $2\pi/g=\pi c/e$. So, if we break a Cooper pair and 
transport one of the electrons
around the vortex, the wave function of the system will acquire a nontrivial 
Aharonov-Bohm factor of $-1$. As shown in refs. \cite{Einarsson}, \cite{Kitaev},
whenever transport of local excitations around each other produces such a nontrivial
factor, the ground state of the system on a torus is degenerate, up to an energy 
splitting decreasing exponentially with the system's linear size.
For the present case, it comes out that the ground state degeneracy on a torus is 
at least four-fold.
We do not reproduce the argument here, as it can be found in the above papers. 
Besides, in our case the vacuum structure admits a semiclassical interpretation,
which allows us to obtain all the requisite results in a different way.

Consider classical vacua of (\ref{H}), i.e. configurations of the 
lowest energy. On a torus, these are: 
\begin{eqnarray}
\A & = & \frac{2\pi}{g} \left( \frac{n_{x} \unitvec_{x}}{L_{x}} +
\frac{n_{y} \unitvec_{y}}{L_{y}} \right) \; , \label{Avac} \\
\psi & = & \psi_{0} \exp \left( -2\pi i n_{x} x /L_{x} - 2\pi i n_{y} y /L_{y}  
\right) \; , \label{psivac} 
\end{eqnarray}
where $n_{x}$ and $n_{y}$ are arbitrary integers, $\unitvec_x$ and
$\unitvec_y$ are unit vectors in the two directions, and $L_x$, $L_y$ are 
the corresponding dimensions of the torus; 
\be
\psi_{0}=(a/2b)^{1/2} \; .
\label{psi0}
\ee
The integers $n_{x}$ and $n_{y}$
are the winding numbers of the configuration: they count 
how many times the phase of $\psi$ winds as one travels along the torus's
noncontractible loops. We consider the case when $L_x$ and $L_y$ are comparable
and assume, for definiteness, that
\be
L_x > L_y \; ,
\label{sizes}
\ee
i.e. that the larger loop of the torus is in the $x$ direction.

Tunneling processes mix the perturbative vacua built near the configurations
(\ref{Avac})--(\ref{psivac}) into linear combinations, $\theta$-vacua, analogous
to those of the four-dimensional QCD \cite{theta}. If we denote the perturbative vacua
as $|n_x, n_y\rangle$, the $\theta$-vacua are
\be
|\theta_x, \theta_y\rangle = \sum_{n_x, n_y} \exp(i\theta_x n_x + i\theta_y n_y) \; ,
\label{theta-vacua}
\ee
where $\theta_x$ and $\theta_y$ run from 0 to $2\pi$.
In this case we need two $\theta$ angles because there are two winding numbers, $n_x$ and
$n_y$. A more important difference from QCD, though, is that in the present case the
tunneling amplitudes, and hence the energy splittings among the $\theta$ vacua, are
exponentially suppressed with $L_x$ or $L_y$. This exponential suppression was found
in ref. \cite{Wen&Niu} in a slightly different context, see also ref. \cite{Kitaev}.
It can be explained as follows. A typical tunneling fluctuation consists of a vortex 
and an antivortex, which travel along topologically distinct routes: the vortex travels
distance $\Delta L$, and the antivortex distance $L_y-\Delta L$ (if we consider transitions
that change $n_x$). At least one of these distances is macroscopically large, and to 
travel that far the object has to move very fast,
or to stay in existence for very long, or to achieve a good balance between these two 
extremes. One finds \cite{Wen&Niu} that even the fluctuation that achieves the optimal 
balance still has a Euclidean action proportional to $L_y$, resulting in
an exponentially suppressed amplitude.

In what follows we will assume that system is sufficiently large, so that the tunneling 
processes that change $n_x$ and $n_y$ are practically nonexistent. In this case, the
linear combinations (\ref{theta-vacua}) are no longer special, and an equally good
basis in the ground state subspace is provided by the perturbative vacua $|n_x, n_y\rangle$
built near the classical solutions (\ref{Avac})--(\ref{psivac}). From the nontrivial
properties of excitations, with respect to transport around each other, we have learned
that, when tunneling is neglected, there are at least four degenerate ground states.
Now we find infinitely many degenerate vacua $|n_x, n_y\rangle$. It is easy to make
four from infinitely many. Note that the ground states $|n_x, n_y\rangle$ and
$|n_x + 1, n_y\rangle$ can be distinguished by breaking a Cooper pair and transporting one
of the electrons around the torus in the $x$ direction. Say, for $n_x=0$ the electron will
pick no phase factor, while for $n_x=1$ it will pick a factor of $-1$. On the other hand, 
given that
the charge of electron is the minimal charge in the system, there is no way to distinguish
between $|n_x, n_y\rangle$ and $|n_x + 2, n_y\rangle$. Similarly, one cannot distinguish between
$|n_x, n_y\rangle$ and $|n_x, n_y + 2\rangle$. So, in the
absence of tunneling, instead of the infinitely many vacua $|n_x, n_y\rangle$ we may as well
consider only four ``equivalence classes'', corresponding to $n_x$ and $n_y$ both being even,
one being even, the other odd, and both being odd, respectively. The four vacua deduced from
the quantum numbers of the excitations are representatives of these four equivalence classes.

Although, as we have seen, in the absence of tunneling we do not have to 
consider the entire infinite ``lattice'' of the vacua $|n_x, n_y\rangle$, sometimes it is
convenient to do so. In particular, in the next section we will
see that thermal fluctuations in the winding numbers are conveniently viewed as diffusion
of $n_x$ and $n_y$ over an infinite lattice made by pairs of integers.

Now consider a type-II film that sits on the surface of a solid torus,
a ``doughnut'', whose bulk is not superconducting. There are still two winding numbers,
$n_x$ and $n_y$. For example, $n_x$ in this case is simply
the total magnetic flux through the doughnut's hole, in units of the flux
quantum $\Phi_0=2\pi/g$.
One can change $n_x$ to $(n_x+1)$ by dragging an extra flux quantum from the outside,
through the bulk of the doughnut. 
This is equivalent
to creating a vortex and an antivortex on the outer side of the doughnut, transporting them 
along topologically distinct paths to the inner side, and annihilating them there,
cf. ref. \cite{Wen&Niu}. In quantum theory, this process occurs spontaneously, as a quantum
fluctuation. It is a tunneling process between two distinct ground states that differ by one 
unit of $n_x$. 
The conclusion that the tunneling rate is suppressed exponentially with $L_y$ (for transitions
that change $n_x$) still applies.

One can switch between the ground states ``by hand'', i.e. by dragging appropriate fluxes
with the help of external solenoids. Switching from $|\psi_1\rangle$ to $|\psi_2\rangle$, 
for a system
that was initially in the linear superposition (\ref{gs}), is equivalent to interchanging
$c_1$ and $c_2$. It is hard to say, though, if this ``quantum
switch'' can serve any useful practical purpose.

On a ``doughnut'', the superconducting current in a ground state with $n_x \sim 1$ and
$n_y = 0$ is of order
$c \Phi_0 /{\cal L}$, where ${\cal L}$ is the self-inductance of the device:
${\cal L} \sim L_x \ln(L_x/L_y)$; in what follows we assume the logarithm here to be of 
order one. The current density
is then inversely proportional to the surface area of the sample. Because the current density
is so low, ground states differing by a few units of $n_x$ or $n_y$
are practically indistinguishable with regard to energies
of local excitations. This circumstance plays an important role in preservation of
quantum coherence at finite temperature.

On the doughnut, as opposed to the ``mathematical'' torus---a rectangle
with periodic boundary conditions, the ground states corresponding to different values of
$n_x$ and $n_y$ are not
exactly degenerate even in classical theory: for different values of $n_x$
there are different amounts of energy associated with the magnetic field 
trapped in the doughnut's hole. 
This energy has very little influence on the rate of topological transitions, so
calculation of the rate can be carried out on the ``mathematical'' torus.
On the other hand,
a real device will be a ``doughnut'', and in that case the magnetic energy
will lead to discrete (labelled by $n_x$)
energy levels. A resonator tuned to the energy difference between two such levels may 
then be able to write linear superpositions of quantum states to this device, or
to a solid superconducting torus, which we discuss later. Estimates related to
this writing technique are given in the concluding section.

\section{Topological transitions at finite temperature}
Consider, for definiteness, 
two ground states that differ by one unit of $n_x$ and have the same $n_y$.
These states may play the role of the basis states $|\psi_1\rangle$ and $|\psi_2\rangle$
forming the linear combination (\ref{gs}). When the system is at a finite
temperature, excited states are also occupied. For low-lying, perturbative excited states, 
we can distinguish between states built near $|\psi_1\rangle$, let us call them
$|\psi_1^{(n)}\rangle$, and those
built near $|\psi_2\rangle$, call them $|\psi_2^{(n)}\rangle$; we neglect tunneling
between these two sectors. 
It is essential that, for superconducting tori that we consider here, excitation energies
are nearly the same for the two sectors, even when (as on a ``doughnut'') there is a
sizable difference between the corresponding ground state energies. If $E_i^{(n)}$ denotes
the energy of $|\psi_i^{(n)}\rangle$, $E_i^{(0)}$ the energy of $|\psi_i\rangle$
($i=1,2$), and the excitation energies are, to a good accuracy, equal:
\begin{equation}
E_1^{(n)} - E_1^{(0)} = E_2^{(n)} - E_2^{(0)} \equiv \Delta E^{(n)} \; ,
\label{exen}
\end{equation}
there is a quasiequilibrium state described by the following density matrix:
\begin{equation}
\rho = \sum_{n} w_{n} |\psi^{(n)}\rangle \langle \psi^{(n)} | \; ,
\label{densmat}
\end{equation}
where
 \begin{equation}
|\psi^{(n)}\rangle = c_{1} |\psi_{1}^{(n)}\rangle + c_{2} 
|\psi_{2}^{(n)}\rangle \; ,
\label{psin}
\end{equation}
and $w_n = {\cal N} \exp(-\Delta E^{(n)} /T)$; ${\cal N}$ is a normalization factor,
and $T$ is temperature. This state is no less quantum-coherent than 
the ground state (\ref{gs}).

The density matrix (\ref{densmat}) is not a fully equilibrium one, 
because it does not include occupation of higher excited states, those that 
cannot be regarded
as being ``near'' either $|\psi_1\rangle$ or $|\psi_2\rangle$. Thermal occupation of these
states will determine the rate of transitions between the two sectors. Our goal in this 
section is to set up formalism for calculating these transition rates. We begin with
the case of a type-II film on the surface of a ``doughnut''.

Any two configurations that differ 
by one unit of $n_{x}$ or one unit of $n_{y}$ are separated by
a potential barrier whose height is, to a good accuracy, twice the energy of a static vortex.
In general, a system at finite temperature does not need to tunnel under a barrier; 
it can go over it as a result of a thermal fluctuation. 
In many cases, the rate of these thermal transitions can be computed by considering vicinity
of the fluctuation corresponding to the top of the barrier \cite{Langer}.
This fluctuation is called the critical fluctuation.
For a toroidal type-II film, however, calculational schemes based on expanding near 
a critical fluctuation 
are completely useless, for the following reason. The top of the barrier in this case
corresponds to a vortex and  an antivortex 
separated by distance $L_y/2$ (for transitions that change 
$n_x$), see Fig. 1. But at a finite
temperature there is a finite density of vortices and antivortices, with a typical distance
between them that is much smaller than $L_y/2$. 
In this situation, a pair of widely separated vortex and antivortex cannot have any special
significance.
Accordingly, we expect that the rate of topological transitions will be determined by 
motion of vortices already populating the medium. Nucleation and 
annihilation of vortex-antivortex pairs will merely 
maintain the equilibrium concentrations of vortices and antivortices.

\begin{figure}
\leavevmode\epsfxsize=3.2in \epsfbox{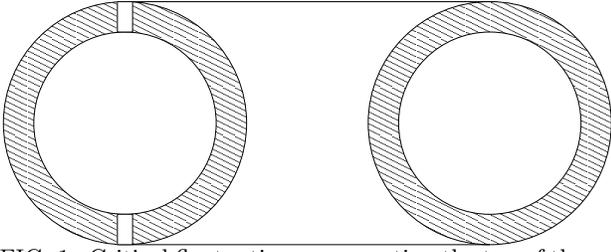}
\caption{Critical fluctuation, representing the top of the potential
barrier between neighboring classical vacua, in the case when 
only the surface of torus is superconducting (shaded area).
The critical fluctuation consists of a vortex and an antivortex lying
on a cross-sectional diameter. It ``melts'' at a finite temperature, because
many vortices intervene between these two, and so plays no role in thermal
transitions between the vacua.}
\label{fig:surface}
\end{figure}

In contrast to the two-dimensional case (film), a critical fluctuation can be readily
identified at finite temperature in a solid superconducting torus, or a loop of thick 
superconducting wire. A solid type-II torus has multiple
ground states, although not as many of them as a torus in which superconductivity is confined to 
the surface. Loops in the $y$ direction are contractible through superconductor, so there is
no winding number $n_y$ that would correspond to those. But $n_x$ still exists
and still counts the number of flux quanta trapped inside the loop.
Changing $n_x$ by dragging a flux through the 
loop is still operational, but instead of a vortex-antivortex pair this procedure now
creates one long Abrikosov flux line through the wire's bulk. The top of
the energy barrier is reached when the flux line is along a 
cross-sectional diameter of the wire, see Fig. 2.
The energy of this critical flux line is $E_0 \propto L_y$. The rate of change in
$n_x$ via thermal fluctuations is proportional to
the Boltzmann factor $\exp(-E_0/T)$ and thus decreases exponentially with $L_y$. 
At zero temperature, when spontaneous topological 
transitions have to be through tunneling, the suppression is even stronger:
a tunneling path is now a worldsheet in the Euclidean spacetime, 
and the tunneling rate goes as an exponential of $L_y^2$. 

\begin{figure}
\leavevmode\epsfxsize=3.2in \epsfbox{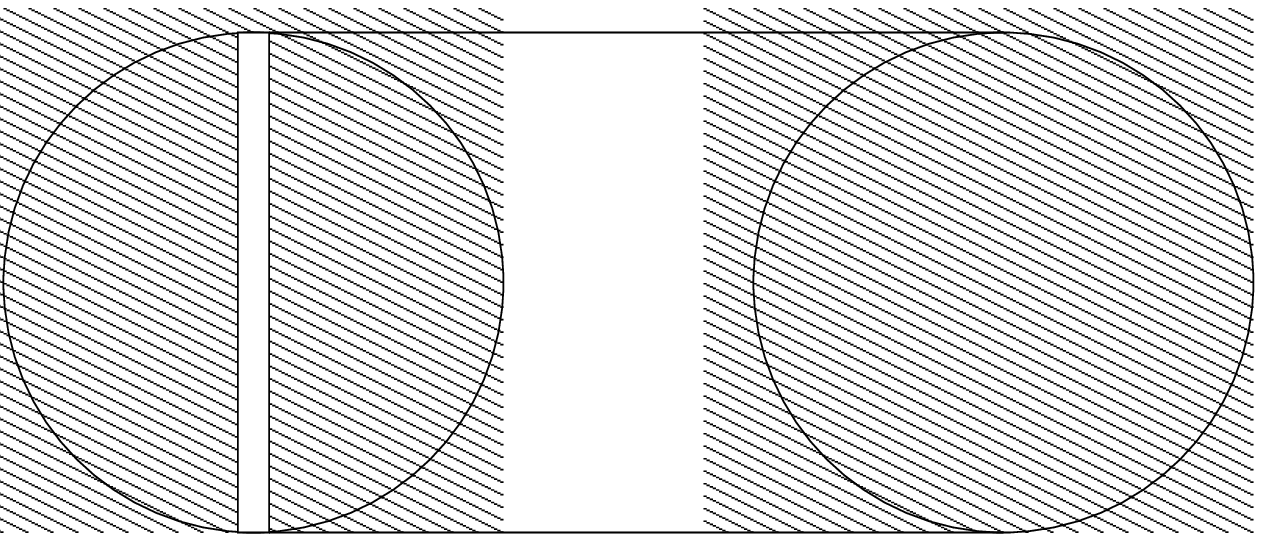}
\caption{Critical fluctuation in a solid type-II superconducting torus. 
Shown is a cross section of the torus. 
The critical fluctuation in this case is an Abrikosov flux
line lying along a cross-sectional diameter.
}
\label{fig:solid}
\end{figure}

What we need for the case of a film
is a definition of the rate of topological transitions that would
make no mention of a critical fluctuation. This requirement is in fact familiar from
studies of topological transitions in the electroweak theory, where depending on the temperature
one may or may not have a critical fluctuation to expand about. A general definition of the rate
in that case is obtained by considering topological transitions as diffusion (or random walk)
of the winding number \cite{KS}. The rate of the transitions is simply the diffusion rate. 
Here we construct a similar definition for toroidal superconducting film.

As we already mentioned, to calculate the rate for the film
it is sufficient to consider the
``mathematical'' torus, on which the classical vacua are given by
(\ref{Avac})--(\ref{psivac}) and are exactly degenerate.
To describe diffusion of the winding numbers, we need to generalize their 
definition so that it will apply away from the vacuum configurations. 
This generalization is not 
unique, but the result for the rate will be the same as long as the 
newly defined winding numbers are equal to $n_{x}$ and $n_{y}$ on
the classical vacua (\ref{Avac})--(\ref{psivac}). A suitable definition is
\begin{eqnarray}
\alpha_{x} & = & \frac{g}{2\pi L_{y}d} \int d^{3} x A_{x} \; , 
\label{alphax} \\
\alpha_{y} & = & \frac{g}{2\pi L_{x}d} \int d^{3} x A_{y} \; .
\label{alphay}
\end{eqnarray}
Note that the winding numbers $\alpha_{x}$ and $\alpha_{y}$ are 
noninteger away from the classical vacua.

Diffusion of $\alpha_{x}$, $\alpha_{y}$ is due to 
diffusive motion of vortices. We assume that the sample is homogeneous
enough so that most of the vortices are not pinned. 
Translational motion of vortices is semiclassical,
so we can define the diffusion rates from the classical 
equilibrium correlator
\begin{equation}
\langle [\alpha_{x}( t) -\alpha_{x}(0)]^{2} \rangle = 2\Gamma_{x} t
\label{bm}
\end{equation}
and a similar one for $\alpha_{y}$. The linear dependence 
on time on the right-hand side is characteristic of diffusion (in the 
absence of external forces), and $\Gamma_{x}$ is the definition of the rate.
Eq. (\ref{bm}) applies at times large compared to some microscopic 
time characterizing interactions of vortices with the heat bath.

The precise meaning of the classical averaging in (\ref{bm}) is as 
follows. For each 
set of initial conditions (for the full fields $\A$ and $\psi$), we 
compute $\alpha_{x}(0)$, then evolve the system
until time $t$, and compute $\alpha_{x}( t)$. The square of the 
difference is then averaged over all 
initial conditions, using the Boltzmann distribution for those. At this 
point, we should remember however that the system (\ref{H}) 
is not isolated but evolves under the influence of a heat bath.
The heat bath is comprised by all degrees of freedom not explicitly
present in (\ref{H})---specifically, those associated with electrons.
So, the requisite evolution equation includes a random (Langevin) force,
and we need to average over realizations of that force as well. 

Before we proceed, it is convenient to recast the definition of the 
rate into a different form, which is more convenient for actual 
calculation. The procedure is completely standard. First,
the left-hand side of (\ref{bm}) is trivially rewritten as
\begin{equation}
 \int _{0}^{t} dt' \int_{0}^{t} dt'' \langle \dot{\alpha}_{x}(t') 
 \dot{\alpha}_{x}(t'') \rangle \; .
 \label{rewr}
 \end{equation}
The correlator of time derivatives in (\ref{rewr}) is an equilibrium 
correlator and thus depends only on the difference $t'-t''$. 
We assume that the 
corresponding correlation time is finite (this assumption can be verified
in our specific case). Then, at large $t$ the integral (\ref{rewr}) is well
approximated by
\begin{equation}
 t \int_{-\infty}^{\infty} d\tau \langle \dot{\alpha}_{x}(\tau) 
 \dot{\alpha}_{x}(0) \rangle \; ,
 \label{rewr2}
 \end{equation}
which allows us to rewrite the definition (\ref{bm}) of the rate 
$\Gamma_{x}$ as
\begin{equation}
\Gamma_{x} = {1\over 2} \int_{-\infty}^{\infty} d\tau
\langle \dot{\alpha}_{x}(\tau)  \dot{\alpha}_{x}(0) \rangle \; .
\label{rate}
\end{equation}
As we will now show, the rate can be found explicitly by a simple 
calculation based on the picture of diffusing vortices.

\section{Calculation of the rate}
When a vortex crosses line $y=b$, the line integral
\begin{equation}
C_{b} = \int_{y=b} A_{x} dx
\label{Cb}
\end{equation}
changes by the amount of the vortex flux, i.e.
\begin{equation}
C_{b} \to C_{b} \pm 2\pi / g \; ,
\label{change}
\end{equation}
the sign depending on which direction the vortex is headed.
If a vortex moves the entire length $L_{y}$ (in the $y$ direction), it
crosses all such lines and changes $\alpha_{x}$, which is essentially
the average of $gC_{b}/2\pi$ over $b$, by $\pm 1$.
So, if a vortex moves a distance $\Delta y$, it changes $\alpha_{x}$
by the amount
\begin{equation}
\Delta \alpha_{x} = \frac{\Delta y}{L_{y}} \; .
\label{dalpha}
\end{equation}
Taking into account all the vortices (of which there are $N_{v}$) and 
antivortices (of which there are $N_{a}$), we then
obtain the time derivative of $\alpha_{x}$ as follows
\begin{equation}
\dot{\alpha}_{x} = \frac{1}{L_{y}}\left(
\sum_{v=1}^{N_{v}} \dot{y}_{v} - \sum_{a=1}^{N_{a}} \dot{y}_{a} 
\right) \; .
\label{sum}
\end{equation}
We now substitute this expression into the formula (\ref{rate}) for 
the rate and assume that, because the vortices are well separated,
the velocities of different vortices are uncorrelated. We obtain
\be
\Gamma_{x} = \frac{N_{v}+N_{a}}{2L_{y}^{2}} \int_{-\infty}^{\infty} d\tau
\langle \dot{y}(\tau)  \dot{y}(0) \rangle \; .
\label{rate2}
\ee

The correlator of velocities in (\ref{rate2}) is computed using the 
equation of motion for a single vortex. We use a simple Langevin 
equation of the form
\be
M \ddot{\r} + \eta \dot{\r} = {\bf f}(t) \; ,
\label{lang}
\ee
where $M$ is the mass of a vortex, $\eta$ in the viscosity coefficient, and $\f(t)$ 
is a random force, which we assume to be Gaussian white noise; 
$\r$ is the position vector of the vortex, $\r= (x, y)$.
The condition of applicability of (\ref{lang}) is that the response of the
electronic subsystem to changes in $\psi$ and $\A$ is local; otherwise, there would
be a nonlocal response kernel instead of the single coefficient $\eta$. 
The response is
local when the mean-free path $l_{\rm tr}$ of the electrons is much smaller than 
the characteristic
length scale from which $\eta$ receives the main contribution. As we 
will see in Appendix,
the latter length scale is the coherence length of the superconductor $\xi$, so the
condition of applicability of (\ref{lang}) is
\be
l_{\rm tr} \ll \xi \; ,
\label{cond}
\ee
i.e. the superconductor should be sufficiently ``dirty''. 

Calculation of $\eta$ had a long history and has eventually been achieved
on the basis of microscopic theory \cite{LO}. It is more or less straightforward, though,
to obtain an {\em estimate}, so we present it here. (We assume that the condition (\ref{cond}) 
is satisfied.) A moving vortex will constantly transfer parts of its kinetic energy 
to the electrons, which they will dissipate in collisions with lattice impurities.
There are two mechanisms of dissipation \cite{LO}. One is Joule heat, which dissipates an 
amount $\sigma E^2$ of energy per unit time per unit volume; here $\sigma$ is the normal
conductivity of the metal, and $E$ is the electric field created by the vortex motion.
The other mechanism is related to response of the electrons to changes in the magnitude of
$\psi$; it dissipates an amount of order 
$a (\partial_t \psi_0)^2 \tau_{\rm tr}$, where
$\tau_{\rm tr}$ is the electronic mean-free time, and $a$ is the parameter from (\ref{H}). 
These two amounts are typically of the same order of
magnitude, except at temperatures close to critical, where the second 
amount is small.
We estimate $E^2$ created by a moving vortex in Appendix. This allows us to estimate $\eta$
from
\be
\eta v^2 \sim \sigma \int d^{3} x E^2 \; ,
\label{eta}
\ee
where $v$ is the vortex speed. The vortex mass $M$ can be estimated from
\be
{1\over 2} M v^2 \sim {1\over 8\pi} \int d^3 x E^2 \; .
\label{mass}
\ee
In Appendix, we find that the integrals in (\ref{eta})--(\ref{mass}) are saturated
at distances $r\sim \xi$ from the vortex center. Curiously, in our final formula 
for the transition rate, $\eta$ and $M$ will appear only via the ratio
\be
\gamma = \eta / M \sim \sigma \; .
\label{ratio}
\ee
Note that this ratio grows with $\sigma$, i.e. it is larger in a purer
metal (which is still ``dirty'', though, in the sense of (\ref{cond})). 
Physically, this is because electrons in a purer metal more readily accept 
energy from a moving vortex.

From (\ref{lang}), it follows that 
\be
\langle \dot{y}(\tau)  \dot{y}(0) \rangle = 
\langle \dot{y}^{2} \rangle \exp(-\gamma|\tau|) \; ,
\label{corr}
\ee
where $\gamma=\eta/M$, and $\langle \dot{y}^{2}\rangle$ can
be determined by equipartition:
\be
{M \over 2}\langle \dot{y}^{2} \rangle = {T \over 2} \; .
\label{equip}
\ee
Assembling the pieces together, we obtain
\be
\Gamma_{x} = \frac{T}{\eta} \frac{N_{v}+N_{a}}{L_{y}^{2}} \; .
\label{rate3}
\ee
A striking feature of this result is that it does not contain any 
volume enhancement: although there is a macroscopic factor of
$(N_{v}+N_{a})$, it is essentially canceled out by the inverse 
powers of $L_{y}$. The physical reason behind this suppression is 
the extremely long time it takes a vortex to circumnavigate the torus:
diffusion through a distance of order $L_{y}$ requires time of order 
$L_{y}^{2}$.

The total number of vortices and antivortices is determined by the
Boltzmann distribution:
\begin{eqnarray}
N_{v} + N_{a} & = & \frac{2V}{(2\pi)^{2}} \int
\exp[-\beta (F_{0} + p^{2}/2M)] d^{2} p \nonumber \\
 & = & \frac{V}{\pi} \e^{-F_{0}/T} MT \; ,
\label{num}
\end{eqnarray}
where $V=L_{x}L_{y}$ is the total 2d volume and
$F_{0}$ is the free energy required to create a vortex. Using
$F_{0}$ instead of the vortex energy takes into 
account thermal population of the vortex's internal states.
Substituting (\ref{num}) into (\ref{rate3}), we finally obtain
\be
\Gamma_{x} = \frac{MT^{2}}{\pi\eta} \frac{L_{x}}{L_{y}} 
\e^{-F_{0}/T} \; .
\label{rate4}
\ee
This is the rate of transitions that change $\alpha_{x}$.
The rate of those that change $\alpha_{y}$ is obtained by 
interchanging $L_{x}$ and $L_{y}$.

\section{Discussion}
As we have already mentioned, for superconducting film the exponential
factor in (\ref{rate4}) can be made practically as small as one 
wishes, because $F_{0}$ grows linearly with the film's thickness. 
So, a thick film on the surface of a torus or, as the limiting case, 
a solid superconducting torus such as shown in Fig. 2 provide
quantum memory that is stable against thermal fluctuations.
We propose the following way to write to and read from this quantum memory.

Because magnetic field trapped in the hole of a superconducting torus
(or of any other shape with a noncontractible loop)
carries energy, the torus behaves
as a giant ``atom'', in the sense that it has a discrete energy spectrum, with
different levels corresponding to different values of $n_x$. We can write the
absolute value of the
energy difference between levels with $n_x = n_1\geq 0$ and $n_x = n_2 > n_1$ as
\be
\hbar \omega = \frac{\hbar^2 c^2}{e^2 R} (n_2^2 - n_1^2) \; ,
\label{omega}
\ee
where $R$ is of order of the linear size of the system (cf. Sect. 2) and
may depend (presumably weakly)
on $n_1$ and $n_2$. (We have restored $\hbar$ in this formula.)
The corresponding electromagnetic wavelength is
\be
\lambda = \frac{2\pi \alpha_{\rm EM} R}{n_2^2 - n_1^2} \; ,
\label{lambda}
\ee
where $\alpha_{\rm EM}$ is the fine-structure constant.
For $R$ of order of a few cm, and $n_{1,2} \sim 1$, the wavelength given by
(\ref{lambda}) is in the millimeter range.
It is possible that one will be able to write a
linear superposition of quantum states to this device by subjecting it to a pulse
of radiation of frequency $\omega$ in a resonant cavity, similarly to how one
induces Rabi precession in atoms. One may be able to read from this
quantum memory by transferring the linear superposition to radiation field in a
high-$Q$ cavity, as was done for atoms in the experiment of ref. \cite{cavity}.
Unlike a single photon in a cavity or an excited state of an atom, the basis states
in our case are macroscopically entangled, so this device will be able to store 
the linear superposition for a much longer time.

If one wants to operate the read 
and write cavities at their principal resonant frequencies and use single-photon
transitions, at least one of the dimensions of each cavity should be of order $\lambda$.
We propose to use, as quantum memory, a loop of superconducting wire,
such that the cross-sectional diameter of the wire is of order $\lambda$, while
the size of the loop itself is large enough for $R$ to be on the order of centimeters. 
Only short arcs of the loop need to
pass through the write and read cavities.
This arrangement corresponds to $L_y \sim \lambda$ in our formulas.
We expect that
the effective size of the interaction region, for interaction between cavity photons
and the wire, is also of order $\lambda$, and hence of the same order as $L_y$.
In this sense, the interaction is nonlocal, so writing time may be not exponentially
large.
At the same time, $L_y$ is still macroscopic, so the rate of thermal transitions
changing $n_x$ is suppressed.

We leave calculation of the rate of topological transitions induced by a radiation
field for future work and turn, briefly, to systems with non-Abelian anyons.
Theoretically, a diverse set of manipulations on degenerate states is available
for some of these systems \cite{Kitaev}.
It has been argued that non-Abelian anyons are realizable as 
excitations of the Pfaffian state \cite{nonab}. 
The latter is a quantum Hall state with a certain type of 
pairing correlation between electrons and is closely related to 
the state proposed in \cite{HR} as a possible
explanation of the experimentally observed \cite{5/2} $\nu=5/2$ 
Hall plateau. With non-Abelian anyons,
nontrivial topology is not required for a sample to have degenerate
ground states. It is sufficient to ``puncture'' the surface of the sample 
with a few localized excitations (vortices). If the typical distance 
$L$ between these vortices is macroscopic, one expects that the
zero temperature tunneling between the ground states is
suppressed exponentially with $L$ \cite{Kitaev,Preskill}. 

At finite temperature, in addition to
those carefully planted vortices there will be a sea of thermally 
excited ones. What will be the rate at which quantum memory 
deteriorates in this case? The exponential Boltzmann factor, 
like the one in (\ref{rate4}), should still be 
present in the rate. In quantum Hall samples
$F_{0}$ will be the larger of the free energy required to create a 
vortex and the free energy required to unpin it from lattice defects. 
Because these systems are intrinsically two-dimensional, one cannot 
increase $F_{0}$ at will. As an estimate of $F_{0}$, 
we can use the value of temperature corresponding to the onset 
of strong temperature dependence of diagonal resistivity. This value can be 
determined experimentally. According to ref. \cite{5/2}, it is 100 mK 
for the $\nu=5/2$ state described in that paper.

The preexponential factor (prefactor) 
in the rate will be determined by motion of thermally
excited vortices around a localized one. By analogy with the results of the
present paper, we expect that a thermal vortex that is initially at distance $\rho$ from 
the localized one will contribute an amount of order $1/\rho^2$ to the prefactor.
Then, the prefactor will be proportional to
\be
\int \rho d\rho / \rho^2 \sim \ln L' \; ,
\label{log}
\ee
where $L'$ is either the linear size of the sample or the distance between
the localized vortices, so the volume enhancement of the rate will be at most 
logarithmic.

While this paper was being completed, we have learned about a recent proposal
\cite{Mooij&al} to use, as a basis for quantum computation, 
current-carrying states in superconducting loops with Josephson junctions.
The authors of ref. \cite{Mooij&al} propose to obtain linear superpositions
of these basis states by modulating 
magnetic fluxes through the loops with pulses of external current.
This technique may work also for the quantum memory device proposed here,
i.e. one may be able to use an external current instead of a resonating
cavity to change $n_x$. We plan to return to analysis
of this possibility elsewhere.

\acknowledgments

The author thanks T. Clark, S. Kivelson, S. Love, P. Muzikar, and M. Stone 
for discussions, and N. Giordano for pointing out ref. \cite{Mooij&al}.
This work was supported in part by the U.S. Department of Energy under Grant 
DE-FG02-91ER40681 (Task B).

\appendix
\section*{Electric field of a moving vortex}
Electric field produced by a moving vortex determines the vortex mass 
$M$ and the viscosity coefficient $\eta$. 
Here we will compute the electric field produced at large distances from 
the vortex core. We will learn in the process that the region away from the core is 
{\em not} where most of the energy associated with the electric field is concentrated.
This precludes us from actually calculating the vortex mass, but we will 
obtain an order of magnitude estimate.

We begin with a collection of formulas describing a {\em static} vortex, 
in notation close to that of ref. \cite{LP}. The magnetic field of the vortex is in the 
$z$ direction.
We consider the extreme type-II case when the penetration depth $\delta$ 
of magnetic field is
much larger than the coherence length $\xi$. For the GL Hamiltonian (\ref{H}), 
\begin{eqnarray}
1/\delta^{2} & = & 2  g^{2} \zeta \psi_{0}^{2} \; , \label{delta} \\
1/\xi^{2} & = & 2a/\zeta \; . \label{xi}
\end{eqnarray}
When distance $r$ from the center of the vortex is much larger than $\xi$,
the magnetic field of a static vortex located at the origin is approximately
\be
B(x, y) = \frac{1}{g \delta^2} K_0(r/\delta) \; ,
\label{B0}
\ee
where $r=(x^2 + y^2)^{1/2}$, and 
$K_0$ is the Macdonald function of the zeroth order. At $r\neq 0$, this magnetic field
satisfies
\be
\delta^2 \nabla^2 B - B = 0 \; .
\label{B1}
\ee
We also recall that at small values of its argument $K_0$ is logarithmic:
$K_0(z) = -\ln z + O(1)$. 

Now suppose the vortex moves through the origin with velocity $\v$, 
which lies in the $x$--$y$ plane.
The rate of change of the magnetic field is
\be
\partial_t B(x, y) = -\v \cdot \nabla B(x, y) \; .
\label{Bdot}
\ee
The changing magnetic field produces an electric field, which is related to 
$\partial_t B$ via one of Maxwell's equations, 
$(\nabla\times \E)_z = - \partial_t B/c$. 
The general solution to this equation in our case is
\be
\E(x, y) =  - c^{-1} (\v \times \unitvec_z) B(x, y) + \nabla f(x, y) \; ,
\label{E0}
\ee
where $f$ is so far an arbitrary function. We fix $f$ from the condition that
$\nabla \cdot \E = 0$. This condition expresses the absence of charge separation
inside the material; we expect it to hold to a good accuracy because charge 
separation in a metal is associated with a large (plasmon) frequency 
gap. Using (\ref{B1}), we then obtain, at large distances from the core,
\begin{eqnarray}
\E(x, y) & = & - c^{-1} (\v \times \unitvec_z) B(x, y) \nonumber \\
 & & \mbox{} + c^{-1} \delta^2 (v_y \partial_x - v_x \partial_y) \nabla B(x, y) \; .
\label{E1}
\end{eqnarray}
At small $r$, the first term here goes as $\ln r$, but the second term goes
as $1/r^2$. When the second term dominates, $E^2 = v^2/(g^2 c^2 r^4)$.

Kinetic energy of the vortex is 
\be
K = {1\over 2} M v^2 \sim {1 \over 8\pi} \int d^3 x E^2 \; .
\label{K}
\ee
For the field (\ref{E1}), the integral in (\ref{K}) diverges at small $r$, 
due to the singular second term in (\ref{E1}). This means that the main contribution to 
the mass comes from the core of the vortex, where (\ref{E1}) does not apply.
Nevertheless, we can obtain an order of magnitude estimate for the mass by using (\ref{E1})
and cutting of the divergence at distances of the order of the core radius, $r\sim \xi$.
This gives
\be
M \sim \frac{d }{e^{2} \xi^2} \sim \frac{H_{c2}d}{ec} \; ,
\label{M}
\ee
where $d$ is the thickness of the film. The second estimate in 
(\ref{M}) uses 
the upper critical field $H_{c2}\sim \Phi_{0}/\xi^{2}$, where 
$\Phi_{0}=\pi c/e$ is the flux quantum. From eq. (\ref{ratio}), we can 
now obtain an estimate for the viscosity coefficient $\eta$:
\be
\frac{\eta}{d} \sim \frac{\sigma H_{c2}}{ec} \; ,
\label{eta1}
\ee
which is in agreement with the results of calculations based on 
microscopic theory \cite{LO}.


\begin{thebibliography}{99}
\bibitem{op} A good order parameter 
for superconductor is penetration depth of magnetic field, which is finite in the
superconducting phase and infinite in the normal phase.
\end{thebibliography}
\end{document}